\documentclass[aps,pre,floats,floatfix,notitlepage]{revtex4-1}

\usepackage{amsmath}

\newcommand{\rmrk}[1]{#1}

\begin{document}

\title{Path integral approach to the quantum fidelity amplitude}

\author{Ji\v{r}\'{\i} Van\'{\i}\v{c}ek$^{1}$, Doron Cohen$^{2}$}

\address{
{$^{1}$Laboratory of Theoretical Physical Chemistry, Institut des Sciences et
Ing\'{e}nierie Chimiques, Ecole Polytechnique F\'{e}d\'{e}rale de Lausanne
(EPFL), CH-1015, Lausanne, Switzerland} \\
{$^{2}$Department of Physics, Ben-Gurion University of the Negev, Beer-Sheva 84105, Israel}
}




\onecolumngrid

\begin{abstract}
The Loschmidt echo is a measure of quantum irreversibility and is determined
by the fidelity amplitude of an imperfect time-reversal protocol.
Fidelity amplitude plays an important role both in the foundations of quantum mechanics
and its applications, such as time-resolved electronic spectroscopy.
We derive an exact path integral formula for the fidelity amplitude and use it
to obtain a series of increasingly accurate semiclassical
approximations by truncating an exact expansion of the path integral exponent.
While the zeroth-order expansion results in a remarkably simple, yet
nontrivial approximation for the fidelity amplitude, the first-order expansion
yields an alternative derivation of the so-called "dephasing
representation", circumventing the use of semiclassical
propagator as in the original derivation. We also obtain approximate
expression for fidelity based on the second-order expansion, which resolves
several shortcomings of the dephasing representation. The rigorous derivation
from the path integral permits the identification of sufficient conditions
under which various approximations obtained become exact.
\end{abstract}

\maketitle

\section{\label{sec:intro}Introduction}

Because of the unitarity of quantum evolution, the overlap of two different
quantum states remains constant in time. As a consequence, to measure
stability of quantum dynamics, one has to perturb the Hamiltonian rather than
the initial state. For this purpose, Peres has introduced \cite{peres:1984}
the notion of \emph{quantum fidelity}, defined for pure initial states $\psi$
as $F\left(  t\right)  :=\left\vert f\left(  t\right)  \right\vert ^{2}$,
where
\begin{equation}
f\left(  t\right)  :=\left\langle \psi\left\vert e^{+iH^{\prime}t/\hbar
}e^{-iH^{\prime\prime}t/\hbar}\right\vert \psi\right\rangle \label{eq:f}%
\end{equation}
is the fidelity amplitude, $H^{\prime}$ is the unperturbed Hamiltonian, and
$H^{\prime\prime}=H^{\prime}+\Delta H$ is the perturbed Hamiltonian. Equation
(\ref{eq:f}) states that fidelity amplitude is the overlap at time $t$ of two
identical initial states evolved with two different time-independent Hamiltonians.

Fidelity is also referred to as the \emph{Loschmidt echo}
\cite{Pastawski_Levstein:2000} because it can be interpreted as the survival
probability of an initial state~$\psi$ evolved for time $t$ with Hamiltonian
$H^{\prime}$ and subsequently for time $-t$ with $H^{\prime\prime}$. It has
been studied extensively in the past fifteen years
\cite{Gorin_Prosen:2006,Jacquod_Petitjean:2009,Goussev_Wisniacki:2012} leading
to the identification of various universal regimes of its decay in time, which
are closely related to similar observations in the theory of wavepacket
dynamics and to the parametric regimes of the local density of states
\cite{Cohen_Heller:2000,Wisniacki_Cohen:2002}.

Quantum fidelity has a fundamental role in our understanding of quantum
irreversibility \cite{Hiller_Kottos:2004}; it provides another perspective to
the theories of decoherence; and it is important for experimental realizations
of quantum computation \cite{nielsen:2000}.
\rmrk{While several nuclear
magnetic resonance \cite{Pastawski_Levstein:1995,Usaj_Pastawski:1998},
microwave \cite{Schafer_Stockmann:2005}, and atom optics
\cite{Andersen_Davidson:2006,Wu_Prentiss:2009} experiments were designed
specifically to study the Loschmidt echo or fidelity amplitude}, the same
correlation function occurs naturally in linear and nonlinear electronic
spectroscopy. For example, within the time-dependent perturbation theory and
Condon approximation, electronic absorption or emission spectra, and
time-resolved spectra in particular, can be computed via a Fourier transform
of an appropriately defined fidelity amplitude
\cite{Mukamel:1982,Wehrle_Sulc:2011,Sulc_Vanicek:2013}.

Loschmidt echo has been studied by many different approaches, which are
reviewed in
Refs.~\cite{Gorin_Prosen:2006,Jacquod_Petitjean:2009,Goussev_Wisniacki:2012}.
Here we focus on a path integral approach, in order to gain further
understanding of the often used semiclassical methods. Indeed, many of the
analytical expressions for fidelity decay were obtained by the original
semiclassical approach of Jalabert and Pastawski
\cite{Jalabert_Pastawski:2001}, while Cerruti and Tomsovic
\cite{Cerruti_Tomsovic:2002} performed the first numerical semiclassical
calculation in which they found explicitly $\sim1000$ stationary-phase
contributions to fidelity amplitude. Van\'{\i}\v{c}ek and Heller
\cite{Vanicek_Heller:2003} avoided the search for stationary-phase points and
obtained a uniform expression for fidelity by combining Miller's initial value
representation \cite{Miller:1970,Miller:2001} with the semiclassical
perturbation approximation \cite{Miller_Smith:1978}. This surprisingly simple
and accurate expression, although limited to wave packets localized in
position, was successfully applied as a starting point to derive the decay of
fidelity in the deep Lyapunov regime \cite{Wang_Li:2005} and the plateau of
fidelity in neutron scattering \cite{Petitjean_Bevilaqua:2007}. By linearizing
the semiclassical initial value representation of the fidelity amplitude,
Van\'{\i}\v{c}ek later obtained \cite{Vanicek:2004,Vanicek:2006} a more
general and accurate approximation, the so-called \emph{dephasing
representation},
\begin{equation}
f_{\text{DR}}(t)\ \ =\ \ \int\frac{d^{2D}x_{0}}{h^{D}}\ \rho_{W}(x_{0}%
)\ \exp\left[  -\frac{i}{\hbar}\int_{0}^{t}\Delta H(x(s))ds\right]  ,
\label{eq:f_DR}%
\end{equation}
applicable not only to pure states ($\rho=|\psi\rangle\langle\psi|$), but also
to arbitrary mixed initial states $\rho$. In Eq.~(\ref{eq:f_DR}), $D$ is the
number of degrees of freedom, $x:=(q,p)$ is a collective notation for
positions $q$ and momenta $p$, $h=2\pi\hbar$ is the Planck constant, $x(t)$
denotes the phase space coordinates at time $t$ of a trajectory of the average
Hamiltonian $H:=(H^{\prime}+H^{\prime\prime})/2$ with initial condition
$x_{0}$, and $\rho_{W}$ is the Wigner function, i.e., the Wigner transform of
the density operator $\rho$ of the initial state. Note that we use the
following convention for the Wigner transform of a general operator $A$:%
\[
A_{W}(x):=\int d^{D}\xi\,\langle q-\xi/2|A|q+\xi/2\rangle e^{ip\cdot\xi/\hbar
}.
\]

In electronic spectroscopy, the dephasing representation and closely related
approximations are known as Mukamel's phase-averaging method
\cite{Mukamel:1982,book_Mukamel} or Wigner-averaged classical limit, and were
used by various authors
\cite{Shemetulskis_Loring:1992,Rost:1995,Egorov_Rabani:1998,Shi_Geva:2005}. In
the context of the mixed quantum-classical Liouville equation, Martens and
coworkers obtained a similar expression for the evolution of coherences of the
density operator \cite{Martens_Fang:1997,Riga_Martens:2006}. In the field of
quantum chaos, the dephasing representation successfully described, e.g., the
local density of states and the transition from the Fermi-Golden-Rule to the
Lyapunov regime of fidelity decay
\cite{Ares_Wisniacki:2009,Wisniacki_Ares:2010,Garcia-Mata_Wisniacki:2011}.

Yet the most attractive feature of the dephasing representation is its
efficiency: Motivated by numerical comparisons with other semiclassical
methods \cite{Wehrle_Sulc:2011}, it was proved analytically
\cite{Mollica_Vanicek:2011} that the number of trajectories required for
convergence of the dephasing representation was independent of the system's
dimensionality, Hamiltonian, or total evolution time. Unlike its efficiency,
the accuracy of the dephasing representation is not always sufficient. This
approximation is exact in displaced harmonic oscillators
\cite{Mukamel:1982,book_Mukamel} and often accurate in chaotic systems
\cite{Vanicek:2004,Vanicek:2006}, but it breaks down in as simple systems as
harmonic oscillators with different force constants. This problem can be
partially remedied by augmenting the approximation with a~prefactor
\cite{Zambrano_Almeida:2011,Zambrano_Vanicek:2013} which, however, is still
not exact even for harmonic systems.

\textbf{Outline.-- } The present paper was motivated by two goals: First, to
derive the dephasing representation from the Feynman path integral, without
employing the semiclassical propagator, and, second, to obtain a semiclassical
approximation correcting the drawbacks of the original version of the
dephasing representation. Below, we do exactly that, but on the way also
obtain a recipe for obtaining increasingly accurate semiclassical
approximations from the expansion of the path integral, and explicit
expressions for the zeroth, first, and second-order expansions. As we will
see, the first-order expansion yields the original dephasing representation,
and its inaccuracies can be corrected with the second-order expansion. The
paper is organized as follows: First, in Sec.~\ref{sec:path_integral} we
derive the coordinate-space path integral representation of fidelity amplitude
by analogy with the path integral for the classical Liouville propagator and
quantum propagator of the density operator. Then, in
Sec.~\ref{sec:phase-space_PI} we provide an alternative and more explicit
phase-space path integral representation of fidelity amplitude in
\rmrk{kicked} quantum maps, which allows us to obtain the zeroth,
first, and second-order approximations. Section~\ref{sec:discussion} discusses
under which circumstances various approximations are exact, while
Sec.~\ref{sec:conclusion} concludes the paper.


\section{\label{sec:path_integral}Coordinate-space path integral
representation}

In order to simplify our first derivation of a path integral representation of
$f(t)$, in this section we will consider one-dimensional systems described by
the Hamiltonian
\begin{equation}
H=\frac{p^{2}}{2m}+V(q). \label{eq:H}%
\end{equation}
The derivation is based on analogies with path integral propagators of
classical and quantum densities, which were discussed in detail by Cohen for
systems with noise in Ref.~\cite{Cohen:1997}.


\subsection{\label{subsec:quantum_prop}Quantum propagator}

\rmrk{The quantum propagator of a wave function
can be obtained from the well known Feynman path-integral expression}
\begin{equation}
U(q|q_{0};t)\ \ :=\ \ \langle q|e^{-iHt/\hbar}|q_{0}\rangle\ \ =\ \ \int%
_{q_{0}}^{q}\!\!\!\!\mathcal{D}q\ \ \exp\left\{  \frac{i}{\hbar}\int_{0}%
^{t}d\tau\left[  \frac{1}{2}m\dot{q}^{2}-V(q)\right]  \right\}  .
\end{equation}
The density operator evolves as $\rho(t)=e^{-iHt/\hbar}\rho(0)e^{iHt/\hbar}$; 
\rmrk{accordingly, its temporal evolution can be expressed
by a propagator $\mathcal{K}$ as}
\begin{equation}
\rho(q^{\prime\prime},q^{\prime};t)\ \ =\ \ \int dq_{0}^{\prime\prime}\int
dq_{0}^{\prime}\,\mathcal{K}(q^{\prime\prime},q^{\prime}|q_{0}^{\prime\prime
},q_{0}^{\prime};t)\,\rho(q_{0}^{\prime\prime},q_{0}^{\prime};0).
\end{equation}
\rmrk{The propagator $\cal{K}$ of the density operator is trivially related to $U$, namely,}
\begin{equation}
\rmrk{\mathcal{K}}(q^{\prime\prime},q^{\prime}|q_{0}^{\prime\prime
},q_{0}^{\prime};t)\ \ =\ \ U(q^{\prime\prime}|q_{0}^{\prime\prime
};t)\ U(q^{\prime}|q_{0}^{\prime};t)^{\ast}.
\end{equation}
\rmrk{Consequently, the path integral expression for $\mathcal{K}$}
involves summation $\mathcal{D}q^{\prime}\mathcal{D}q^{\prime\prime}$ over the
pair of paths $q^{\prime}(\tau)$ and $q^{\prime\prime}(\tau)$. Alternatively,
we may also use the average and difference coordinates $q:=(q^{\prime}%
{+}q^{\prime\prime})/2$ and $r:=q^{\prime\prime}{-}q^{\prime}$; thus the
summation will be $\mathcal{D}q\mathcal{D}r$, namely
\begin{equation}
\rmrk{\mathcal{K}}(q,r|q_{0},r_{0};t)\ =\ \int_{q_{0}}%
^{q}\mathcal{D}q\int_{r_{0}}^{r}\mathcal{D}r\ \ \exp\left[  \frac{i}{\hbar
}\left(  \int_{0}^{t}d\tau m\dot{q}\dot{r}-\int_{0}^{t}d\tau\left[  V\left(
q+\frac{r}{2}\right)  -V\left(  q-\frac{r}{2}\right)  \right]  \right)
\right]  .
\end{equation}
As a final step we would like to transform the quantum propagator to the
Wigner representation. Recall that $\rho_{W}(q,p)$ is the Fourier transform of
$\rho(q,r)$ in the $r\mapsto p$ coordinate. It follows that
\begin{equation}
\rmrk{\mathcal{K}_{W}}(q,p|q_{0},p_{0};t)\ =\ \int_{q_{0},p_{0}}^{q,p}\mathcal{D}%
q\int\mathcal{D}r\ \ \exp\left(  \frac{i}{\hbar}S[q,r]\right)  .
\end{equation}
The integration $\mathcal{D}r$ in the latter expression is not restricted at
the end-points, whereas the integration $\mathcal{D}q$ is restricted at the
end-points both in $q$ and $\dot{q}$. The restriction on $\dot{q}$ at the
end-points is implicit, through the relation $\dot{q}=p/m$. We have used the
notation
\begin{equation}
S[q,r]\ \ =\ \ S_{\text{free}}[q,r]-\int_{0}^{t}d\tau\left[  V\left(
q+\frac{r}{2}\right)  -V\left(  q-\frac{r}{2}\right)  \right]  \label{eq:S_qm}%
\end{equation}
where
\begin{equation}
S_{\text{free}}[q,r]\ \ =\ \ [m\dot{q}(0)r(0)-m\dot{q}(t)r(t)]+\int_{0}%
^{t}d\tau m\dot{q}\,\dot{r}\ \ =\ \ -\int_{0}^{t}d\tau m\ddot{q}%
\,{r.}\label{eq:S_free}%
\end{equation}
In the next subsection we clarify that the leading order estimate of the
quantum propagator leads to the expected classical result.


\subsection{Classical propagator}

The time evolution of a classical phase-space density \rmrk{$\rho%
_{\text{cl}}(q,p;t)$}, under the dynamics that is generated by a classical
Hamiltonian (\ref{eq:H}), is given by the so-called Liouville propagator. For
an infinitesimal time $d\tau$ the explicit expression \rmrk{for the Liouville propagator} is
\begin{equation}
\rmrk{\mathcal{K}_{\text{cl}}}(q_{2},p_{2}|q_{1},p_{1};d\tau)=2\pi
\hbar\,\delta\left(  p_{2}{-}p_{1}+\frac{\partial V}{\partial q}d\tau\right)
\cdot\delta\left(  q_{2}-q_{1}-\frac{p}{m}d\tau\right)  .
\end{equation}
Here a dummy parameter $\hbar$ has been inserted, which cancels with the
phase-space measure $dqdp/(2\pi\hbar)$. Its value does not have any effect
here, but the use of $\hbar$ will make a later comparison to the quantum
mechanical version more transparent. The inverse Fourier-transformed
($p\mapsto r$) version, \rmrk{$\tilde{\rho}_{\text{cl}}(q,r;t)$}, of phase-space representation
\rmrk{$\rho_{\text{cl}}(q,p;t)$} is analogous to the coordinate-space
representation \rmrk{$\rho(q,r;t)$} of the quantum density matrix. \rmrk{(Tilde will be used on classical densities and propagators in coordinate representation, i.e., if their arguments are $q$ and $r$, or $q^{\prime}$ and $q^{\prime\prime}$.)} The associated
Fourier-transformed version of the classical Liouville propagator is
accordingly
\begin{equation}
\rmrk{\tilde{\mathcal{K}}_{\text{cl}}}(q_{2},r_{2}|q_{1},r_{1};d\tau
)=\frac{m}{d\tau}\exp\left\{  \frac{i}{\hbar}\left[  m\left(  \frac{q_{2}%
{-}q_{1}}{d\tau}\right)  (r_{2}{-}r_{1})\ -\ \left(  \frac{r_{1}+r_{2}}%
{2}\right)  \frac{\partial V}{\partial q}d\tau\right]  \right\}  .
\end{equation}
For a finite time, the convolved propagator may be written as a functional
integral
\begin{equation}
\rmrk{\tilde{\mathcal{K}}_{\text{cl}}}(q,r|q_{0},r_{0};t)=\int_{q_{0}}%
^{q}\mathcal{D}q\int_{r_{0}}^{r}\mathcal{D}r\ \ \ \exp\left[  \frac{i}{\hbar
}\left(  \int_{0}^{t}d\tau\,m\dot{q}\dot{r}-\int_{0}^{t}d\tau\,r\,\frac
{\partial V}{\partial q}\right)  \right]  .
\end{equation}
Transforming back to the phase-space variables we get
\begin{equation}
\rmrk{\mathcal{K}_{\text{cl}}}(q,p|q_{0},p_{0};t)\ =\ \int_{q_{0},p_{0}%
}^{q,p}\mathcal{D}q\int\mathcal{D}r\ \ \exp\left(  \frac{i}{\hbar}%
S_{\text{cl}}[q,r]\right)  ,
\end{equation}
where the classical action is
\begin{equation}
S_{\text{cl}}[q,r]\ \ =\ \ S_{\text{free}}[q,r]-\int_{0}^{t}d\tau
\,r\,\frac{\partial V}{\partial q}.\label{eq:S_cl}%
\end{equation}
Note that the classical action is the same as the leading order $r$~expansion
of the quantum action (\ref{eq:S_qm}).


\subsection{Fidelity amplitude}

Now we use the same procedure to obtain an expression for the quantum fidelity
amplitude at time $t$ assuming that the initial preparation is described by
the density matrix $\rho(q^{\prime\prime},q^{\prime})$, and the two
Hamiltonians differ only in their potential energies $V^{\prime}(q)$ and
$V^{\prime\prime}(q)$. The following is the exact Feynman path integral with
unrestricted integration over all possible paths:
\begin{align}
f(t) &  :=\operatorname{Tr}\left(  e^{-iH^{\prime\prime}t/\hbar}\rho
e^{iH^{\prime}t/\hbar}\right)  \\
&  =\int dq\iint dq_{0}^{\prime}dq_{0}^{\prime\prime}\ \rho(q_{0}%
^{\prime\prime},q_{0}^{\prime})\int_{q_{0}^{\prime}}^{q}\mathcal{D}q^{\prime
}\int_{q_{0}^{\prime\prime}}^{q}\mathcal{D}q^{\prime\prime}\ \exp\left[
\frac{i}{\hbar}\left(  S^{\prime\prime}[q^{\prime\prime}]-S^{\prime}%
[q^{\prime}]\right)  \right]  \\
&  =\int dq\iint dq_{0}dr_{0}\ \rho(q_{0},r_{0})\int_{q_{0}}^{q}%
\mathcal{D}q\int_{r_{0}}^{r}\mathcal{D}r\ \exp\left[  \frac{i}{\hbar}\left(
S^{\prime\prime}\left[  q+\frac{r}{2}\right]  -S^{\prime}\left[  q-\frac{r}%
{2}\right]  \right)  \right]
\end{align}
where the single primed quantities such as $S^{\prime}$ correspond to the
evolution with $H^{\prime}$ and the double primed quantities such as
$S^{\prime\prime}$ correspond to $H^{\prime\prime}$. We now use exactly the
same manipulations as in subsection~\ref{subsec:quantum_prop} and write this
expression using phase-space variables:
\begin{equation}
f(t)\ \ =\ \ \int dq\iint dq_{0}dr_{0}\ \rho(q_{0},r_{0})\int_{q_{0},p_{0}%
}^{q,p}\mathcal{D}q\int\mathcal{D}r\ \exp\left[  \frac{i}{\hbar}\Delta
S[q,r]\right]  ,
\end{equation}
where
\begin{equation}
\Delta S[q,r]\ \ =\ \ S_{\text{free}}[q,r]-\int_{0}^{t}d\tau\left[
V^{\prime\prime}\left(  q+\frac{r}{2}\right)  -V^{\prime}\left(  q-\frac{r}%
{2}\right)  \right]  .
\end{equation}
This expression is in one-to-one correspondence with (\ref{eq:S_qm}); so far,
no approximations were involved. The next step is to expand in $r$,
\rmrk{namely}
\begin{equation}
V^{\prime\prime}\left(  q+\frac{r}{2}\right)  -V^{\prime}\left(  q-\frac{r}%
{2}\right)  \ \ \approx\ \ V^{\prime\prime}(q)-V^{\prime}(q)+r\frac{\partial
V}{\partial q}\ \ =\ \ \Delta V(q)+r\frac{\partial V}{\partial q}(q),
\end{equation}
\rmrk{where $V:=(V^{\prime}+V^{\prime\prime})/2$.} Recall that in
the calculation of the quantum propagator, this linear approximation merely
led to the classical propagator since $\Delta V(q)$ was zero. Here we shall
see that the linearization leads to non-trivial quantum results. Notice that
the approximated action, including the \textquotedblleft
free\textquotedblright\ action of (\ref{eq:S_free}), is linear in the
$r(\tau)$ variables. Also it is possible to express $\rho(q_{0},r_{0})$ as a
Fourier integral over $\rho_{W}(q_{0},p_{0})$, involving $\exp(ip_{0}%
r_{0}/\hbar)$. So now all the $r(\tau)$ including $r_{0}$ appear in a linear
fashion in the exponent. Consequently the unrestricted $\mathcal{D}r$
integration, including the $dr_{0}$ integration, results in a product of delta
functions. Subsequently the $\mathcal{D}q$ integration, including the final
$dq$ integration, picks up only the classical trajectories $q_{\text{cl}}%
(\tau)$. We are left with the following very simple approximation:
\begin{equation}
f(t)\ \ \approx\ \ \iint\frac{dq_{0}dp_{0}}{h}\ \rho_{W}(q_{0},p_{0}%
)\ \exp\left[  -\frac{i}{\hbar}\int_{0}^{t}d\tau\,\Delta V(q_{\text{cl}}%
(\tau))\right]  \label{eq:ft}%
\end{equation}
which coincides with the dephasing representation (\ref{eq:f_DR}).


\section{\label{sec:phase-space_PI}Phase-space path integral representation}

In this section, we will use a phase-space path integral approach and
generalize the analysis of the previous section by considering a system with
$D$ degrees of freedom described by the separable Hamiltonian
\begin{equation}
H(x):=T(p)+V(q),
\end{equation}
where $T(p)$ and $V(q)$ are arbitrary functions describing the kinetic and
potential energies.


\subsection{Quantum propagator}

For short times $\tau$, the quantum evolution operator $U(\tau):=\exp
(-iH\tau/\hbar)$ corresponding to Hamiltonian $H$ can be approximated as%
\begin{equation}
e^{-i\tau H/\hbar}=e^{-i\tau T/\hbar}e^{-i\tau V/\hbar}+\mathcal{O}(\tau^{2}).
\label{eq:Lie-Trotter}%
\end{equation}
In order to avoid the questions of convergence of the path integral and to
make our derivations rigorously exact for as long as possible, we will
consider \rmrk{kicked} quantum maps, in which the error term in the
factorization (\ref{eq:Lie-Trotter}) is zero by definition. In other words, in
a \rmrk{kicked} quantum map, the evolution operator for a single
time step is \emph{defined} to be%
\begin{equation}
U:=e^{-i\tau V/\hbar}e^{-i\tau T/\hbar}\text{.} \label{eq:U}%
\end{equation}

The quantum propagator from position $q_{n}$ to $q_{n+1}$ in a single time
step of the map,
\begin{equation}
U(q_{n+1},q_{n};\tau):=\left\langle q_{n+1}\right\vert U\left\vert
q_{n}\right\rangle =h^{-D}\int d^{D}p_{n}\exp\left\{  i\left[  p_{n}\left(
q_{n+1}-q_{n}\right)  -\tau H(q_{n+1},p_{n})\right]  /\hbar\right\}  ,
\label{eq:K_1step}%
\end{equation}
is obtained by inserting the resolution of identity $\operatorname{Id}=\int
dp_{n}|p_{n}\rangle\langle p_{n}|$ between the potential and kinetic evolution
operators in (\ref{eq:U}). By concatenating $N$ single-step propagators, one
finds the propagator from $q_{0}$ to $q_{N}$ in time $t=N\tau$:%
\begin{align}
U(q_{N},q_{0};N\tau)  &  =\left\langle q_{N}\left\vert U^{N}\right\vert
q_{0}\right\rangle =\int\prod_{n=1}^{N-1}d^{D}q_{n}\prod_{j=0}^{N-1}%
\frac{d^{D}p_{j}}{h^{D}}e^{iS_{N}/\hbar},\label{eq:K_Nsteps}\\
S_{N}  &  :=\sum_{n=0}^{N-1}\left[  p_{n}\left(  q_{n+1}-q_{n}\right)  -\tau
H(q_{n+1},p_{n})\right]  , \label{eq:S_Nsteps}%
\end{align}
where $q_{n}$ and $p_{n}$ denote the positions and momenta after $n$ steps. An
appealing feature of the phase-space path integral is the absence of a
complicated prefactor; one only has to consistently use the standard
phase-space measure $d^{D}qd^{D}p/h^{D}$.


\subsection{Fidelity amplitude}

To find the path integral representation of fidelity amplitude (\ref{eq:f}),
we first express $f(t)$ in terms of the quantum propagators:%
\begin{align}
f(t)  &  =\left\langle \psi\left\vert (U^{\prime})^{-N}(U^{\prime\prime}%
)^{N}\right\vert \psi\right\rangle =\operatorname{Tr}\left[  (U^{\prime\prime
})^{N}\rho(U^{\prime})^{-N}\right] \nonumber\\
&  =\int d^{D}q_{0}^{\prime}d^{D}q_{0}^{\prime\prime}d^{D}q_{N}^{\prime}%
d^{D}q_{N}^{\prime\prime}\,U^{\prime\prime}(q_{N}^{\prime\prime},q_{0}%
^{\prime\prime};N\tau)\rho(q_{0}^{\prime\prime},q_{0}^{\prime})U^{\prime
}(q_{N}^{\prime},q_{0}^{\prime};N\tau)^{\ast}\delta(q_{N}^{\prime\prime}%
-q_{N}^{\prime}), \label{eq:f_prop}%
\end{align}
where the single-primed quantities such as $U^{\prime}$ again correspond to
$H^{\prime}$ and double primed quantities such as $U^{\prime\prime}$ to
$H^{\prime\prime}$. By having expressed fidelity amplitude as a trace of the
evolved density $\rho$, all our derivations below remain valid for general
mixed states. After substituting the path integral expression
(\ref{eq:K_Nsteps}) for the two propagators, we get%
\begin{equation}
f(t)=\int d^{D}q_{N}^{\prime}d^{D}q_{N}^{\prime\prime}\prod_{n=0}^{N-1}%
\frac{d^{2D}x_{n}^{\prime}}{h^{D}}\frac{d^{2D}x_{n}^{\prime\prime}}{h^{D}}%
\rho(q_{0}^{\prime\prime},q_{0}^{\prime})\exp[i\left(  S_{N}^{\prime\prime
}-S_{N}^{\prime}\right)  /\hbar]\delta(q_{N}^{\prime\prime}-q_{N}^{\prime}).
\label{eq:f_PI_1}%
\end{equation}
Now it is convenient to change the independent integration variables to the
average and difference coordinates $x:=(x^{\prime}+x^{\prime\prime})/2$ and
$\Delta x:=x^{\prime\prime}-x^{\prime}$,
\begin{align}
f(t)  &  =\int d^{D}\Delta q_{N}\prod_{n=0}^{N}\frac{d^{2D}x_{n}}{h^{D}}%
\prod_{j=0}^{N-1}\frac{d^{2D}\Delta x_{j}}{h^{D}}\rho(q_{0}^{\prime\prime
},q_{0}^{\prime})e^{iA_{N}/\hbar},\label{eq:f_PI_2}\\
A_{N}  &  :=S_{N}^{\prime\prime}-S_{N}^{\prime}-p_{N}\Delta q_{N},\nonumber
\end{align}
where we have also expressed the delta function $\delta(\Delta q_{N})$ in
terms of an integral over a new variable $p_{N}$. After substituting the
$N$-step action (\ref{eq:S_Nsteps}) for $S_{N}^{\prime}$ and $S_{N}%
^{\prime\prime}$ and simplification, one obtains an explicit expression for
the phase,%
\begin{equation}
A_{N}=-\tau\sum_{n=0}^{N-1}\left[  H^{\prime\prime}(q_{n+1}^{\prime\prime
},p_{n}^{\prime\prime})-H^{\prime}(q_{n+1}^{\prime},p_{n}^{\prime})\right]
+\sum_{j=0}^{N-1}(q_{j+1}-q_{j})\Delta p_{j}-\sum_{k=1}^{N}(p_{k}%
-p_{k-1})\Delta q_{k}-p_{0}\Delta q_{0}. \label{eq:A}%
\end{equation}
Note that expression (\ref{eq:f_PI_2}) with (\ref{eq:A}) is \emph{exact} for
\rmrk{kicked} quantum maps even for finite $N$.


\subsection{Expansion of the path integral}

The explicit expressions above in terms of the average and difference
trajectories $x_{n}$ and $\Delta x_{n}$ will now pay off because we can make
increasingly more accurate expansions of the difference $H^{\prime\prime
}(x^{\prime\prime})-H^{\prime}(x^{\prime})$ in powers of $\Delta x$ which is
the only term in the exponent $A_{N}$ preventing us from performing the path
integral (\ref{eq:f_PI_2})\ analytically. This expansion must be done with
care since \emph{both} the trajectory and Hamiltonian change. Let us start
with the full expansion, which is guaranteed to be exact if both $H^{\prime}$
and $H^{\prime\prime}$ have Taylor series that converge on the entire phase
space:%
\begin{align}
H^{\prime\prime}(x^{\prime\prime})-H^{\prime}(x^{\prime})  &  =H^{\prime
\prime}(x^{\prime\prime})-H^{\prime\prime}(x)+H^{\prime\prime}(x)-H^{\prime
}(x)+H^{\prime}(x)-H^{\prime}(x^{\prime})\nonumber\\
&  =\sum_{n=0}^{\infty}\frac{1}{n!}\frac{\partial^{n}H^{\prime\prime}%
(x)}{\partial x^{n}}\left(  \frac{\Delta x}{2}\right)  ^{n}+\Delta
H(x)+\sum_{n=0}^{\infty}\frac{1}{n!}\frac{\partial^{n}H^{\prime}(x)}{\partial
x^{n}}\left(  -\frac{\Delta x}{2}\right)  ^{n}\nonumber\\
&  =\sum_{n=0}^{\infty}\frac{1}{2^{2n}}\left[  \frac{1}{(2n)!}\frac
{\partial^{2n}\Delta H(x)}{\partial x^{2n}}(\Delta x)^{2n}+\frac{1}%
{(2n+1)!}\frac{\partial^{2n+1}H(x)}{\partial x^{2n+1}}(\Delta x)^{2n+1}%
\right]  , \label{eq:DeltaH_expansion}%
\end{align}
where we have introduced the \emph{average Hamiltonian }$H:=(H^{\prime
}+H^{\prime\prime})/2$ and the difference Hamiltonian (or \emph{perturbation})
$\Delta H:=H^{\prime\prime}-H^{\prime}$. Note that for simplicity, we have for
the moment used one-dimensional notation, and moreover, since both $H^{\prime
}$ and $H^{\prime\prime}$ are separable in coordinates and momenta, so are $H$
and $\Delta H$, and expressions such as $(\partial^{n}H(x)/\partial
x^{n})\left(  \Delta x\right)  ^{n}$ stand for $(\partial^{n}T(p)/\partial
p^{n})\left(  \Delta p\right)  ^{n}+(\partial^{n}V(q)/\partial q^{n})\left(
\Delta q\right)  ^{n}$, etc. There are two important observations to make:

First, in the $\Delta x$ expansion (\ref{eq:DeltaH_expansion}), derivatives of
the average Hamiltonian $H$ appear only with the \emph{odd} powers of $\Delta
x$ and derivatives of the perturbation $\Delta H$ appear only with the
\emph{even} powers of $\Delta x$. Second, the average Hamiltonian appears
naturally and plays a prominent role.
\rmrk{The average Hamiltonian must be
used in order to preserve the order of the expansion. Otherwise, e.g., if}
$H^{\prime}$ \rmrk{were used as a reference in displaced harmonic
oscillators, what appears to be a first-order expansion in} $\Delta x$
\rmrk{would in fact be of second order. This has a consequence, explained below in
Sec.~\ref{sec:discussion}, that in displaced harmonic oscillators, the
dephasing representation~(\ref{eq:f_DR}) mentioned in the introduction is exact if the average Hamiltonian}
$H$ \rmrk{is used as reference, but not if} $H^{\prime}$ \rmrk{is used instead [see Eq.(\ref{eq:expansion_H_about_H'})].}

It turns out to be useful to truncate expansion (\ref{eq:DeltaH_expansion}) at
increasing powers of $\Delta x$. As we will see below, both the zeroth and
first-order expansions yield simple analytical results, the latter agreeing
exactly with the dephasing representation. The second-order expansion cannot
be solved fully analytically, but nevertheless yields an appealing extension
of the dephasing representation.


\subsection{Zeroth-order expansion}

Truncating expansion (\ref{eq:DeltaH_expansion}) at the zeroth power of
$\Delta x$, i.e., setting
\begin{equation}
H^{\prime\prime}(x^{\prime\prime})-H^{\prime}(x^{\prime})\approx\Delta H(x),
\label{eq:DH_0}%
\end{equation}
permits an analytical evaluation of almost all integrals in Eq.
(\ref{eq:f_PI_2}) since they involve either exponentials or delta functions.
The result is the zeroth-order approximation of fidelity amplitude,%
\begin{align}
f^{(0)}(t)  &  =h^{-D}\int d^{D}\Delta q_{0}d^{2D}x_{0}\prod_{n=1}^{N}%
d^{2D}x_{n}\delta(x_{n}-x_{n-1})\nonumber\\
&  ~~\times\rho(q_{0}^{\prime\prime},q_{0}^{\prime})\exp\left\{  -\frac
{i}{\hbar}\left[  p_{0}\Delta q_{0}+\tau\sum_{j=0}^{N-1}\Delta H(q_{j+1}%
,p_{j})\right]  \right\} \nonumber\\
&  =h^{-D}\int d^{2D}x_{0}\rho_{W}(x_{0})e^{-it\Delta H(x_{0})/\hbar
}=\left\langle e^{-it\Delta H(x)/\hbar}\right\rangle _{\rho_{W}(x)},
\label{eq:f^0}%
\end{align}
where $t:=N\tau$ and the last expression employs the notation%
\begin{align}
\left\langle A(x)\right\rangle _{\rho(x)}:=h^{-D}\int d^{2D}x\rho(x)A(x)
\end{align}
for a phase-space \textquotedblleft average\textquotedblright\ of a quantity
$A$ weighted by a normalized quasi-probability distribution $\rho$.
Normalization means that $h^{-D}\int\rho(x)d^{2D}x=1$, which is true for the
Wigner function $\rho_{W}$.

Note that in Eq.~(\ref{eq:f^0}) we have obtained a new approximation for
quantum fidelity amplitude---one that is cruder than the dephasing
representation (\ref{eq:f_DR}) and does not even require running trajectories!

Although approximation $f^{(0)}$ only requires phase-space sampling of the
perturbation at initial time, in general it yields a time-dependent fidelity
amplitude. If one replaces $\rho_{W}$ by the classical Boltzmann distribution,
the zeroth-order approximation for fidelity amplitude coincides with an
approximation used for calculations of inhomogeneously broadened spectra and
known as the static classical limit \cite{Egorov_Rabani:1998,Shi_Geva:2005}.

\emph{Example}: A sufficient condition for the zeroth-order approximation
(\ref{eq:f^0}) for fidelity amplitude to be exact is that the zeroth-order
expansion (\ref{eq:DH_0}) itself is exact, which requires the average and
difference Hamiltonians to be of the form $H=\alpha$ and $\Delta
H=\Delta\alpha+\Delta\beta\cdot q+\Delta\gamma\cdot p$, where $\alpha^{\prime
}$, $\alpha^{\prime\prime}$, $\Delta\beta$, and $\Delta\gamma$ are constants,
implying that the original Hamiltonians must be $H^{\prime}=\alpha^{\prime
}-\frac{1}{2}\Delta\beta\cdot q-\frac{1}{2}\Delta\gamma\cdot p$ and
$H^{\prime\prime}=\alpha^{\prime\prime}+\frac{1}{2}\Delta\beta\cdot q+\frac
{1}{2}\Delta\gamma\cdot p$. Corresponding classical motions are linear
growth\ (or decrease) with time of phase space coordinates for $H^{\prime}$,
$H^{\prime\prime}$, and no motion at all for the average Hamiltonian $H$.
Under such conditions, the zeroth-order approximation $f^{(0)}(t)$ is exact
for arbitrary initial states $\rho$.

This can be verified independently by first expressing fidelity amplitude as%
\begin{equation}
f(t)=\operatorname{Tr}\left[  \rho E(t)\right]  , \label{eq:f_echo}%
\end{equation}
in terms of the echo operator%
\begin{equation}
E(t):=U^{\prime}(t)^{\dag}U^{\prime\prime}(t), \label{eq:echo}%
\end{equation}
then using the phase-space representation of the trace in Eq.~(\ref{eq:f_echo}%
),
\begin{equation}
f(t)=h^{-D}\int d^{2D}x\rho_{W}(x)E_{W}(x,t)=\left\langle E_{W}%
(x,t)\right\rangle _{\rho_{W}(x)}, \label{eq:f_PS}%
\end{equation}
and finally evaluating explicitly the Wigner transform of the echo operator
(\ref{eq:echo}), which, after some algebra, in this case turns out to be
$E_{W}(x,t)=\exp\left[  -it\Delta H(x)/\hbar\right]  $, in agreement with
Eq.~(\ref{eq:f^0}).

Incidentally, the above sufficient condition is not necessary. E.g., for
$\Delta H=0$, expression (\ref{eq:f^0}) is trivially exact, $f^{(0)}(t)=1$,
for arbitrary $H$ even though one neglects the nonvanishing higher order terms
of the average Hamiltonian $H$ in expansion (\ref{eq:DeltaH_expansion}).


\subsection{First-order expansion}

The first-order expansion of (\ref{eq:DeltaH_expansion}) approximates the
Hamiltonian difference as%
\begin{equation}
H^{\prime\prime}(x^{\prime\prime})-H^{\prime}(x^{\prime})\approx\Delta
H(x)+\frac{\partial T}{\partial p}\cdot\Delta p+\frac{\partial V}{\partial
q}\cdot\Delta q. \label{eq:DH_1}%
\end{equation}
Again, most integrals can be solved analytically and one obtains, without any
other approximation,%
\begin{align}
f^{(1)}(t)  &  =h^{-D}\int d^{D}\Delta q_{0}d^{2D}x_{0}\prod_{n=1}^{N}%
d^{2D}x_{n}\delta\left(  q_{n}-q_{n-1}-\tau\frac{\partial T}{\partial
p}(p_{n-1})\right) \nonumber\\
&  ~~\times\delta\left(  p_{n}-p_{n-1}+\tau\frac{\partial V}{\partial q}%
(q_{n})\right)  \rho(q_{0}^{\prime\prime},q_{0}^{\prime})\nonumber\\
&  ~~\times\exp\left\{  -\frac{i}{\hbar}\left[  p_{0}\Delta q_{0}+\tau
\sum_{j=0}^{N-1}\Delta H(q_{j+1},p_{j})\right]  \right\} \nonumber\\
&  =h^{-D}\int d^{2D}x_{0}\rho_{W}\left(  x_{0}\right)  \exp\left[  -\frac
{i}{\hbar}\tau\sum_{j=0}^{N-1}\Delta H(q_{j+1},p_{j})\right]  \label{eq:f^1}%
\end{align}
where $q_{n}$ and $p_{n}$ appearing as arguments of $\Delta H$ in the last
expression are no longer independent path integral variables; instead, they
are the uniquely defined position and momentum coordinates of a trajectory
starting at $x_{0}$ after $n$ steps of the classical symplectic map given by
the average Hamiltonian $H$ and corresponding to the quantum map (\ref{eq:U});
these trajectories are given by the recursive relations between $q_{n}$,
$p_{n}$ and $q_{n-1}$, $p_{n-1}$ expressed by the delta functions in the
preceding equation.

To return from quantum maps to continuous Hamiltonian systems, one takes the
limits $\tau\rightarrow0$ and $N\rightarrow\infty$, so that $N\tau=t$ is
constant, obtaining
\begin{align}
f^{(1)}(t)  &  =h^{-D}\int d^{2D}x_{0}\rho_{W}(x_{0})\exp\left[  -\frac
{i}{\hbar}\int_{0}^{t}\Delta H(x(s))ds\right] \label{eq:f^1_DR}\\
&  =\left\langle \exp\left[  -\frac{i}{\hbar}\int_{0}^{t}\Delta
H(x(s))ds\right]  \right\rangle _{\rho_{W}(x_{0})}=f_{\text{DR}}(t).\nonumber
\end{align}
As promised, by using the first-order expansion of $H^{\prime\prime}%
-H^{\prime}$ in the path integral representation of quantum fidelity, we have
obtained exactly the dephasing representation (\ref{eq:f_DR}). On one hand,
this may seem remarkable, since we did not explicitly employ the semiclassical
propagator which had been used in the original derivation of the dephasing
representation \cite{Vanicek:2006}. On the other hand, the semiclassical
propagator can be obtained by a quadratic expansion of the Feynman path
integral propagator, and since we used a linearization of the path integral,
we implicitly went beyond the semiclassical approximation since, in contrast
to usual semiclassical approximations, expression (\ref{eq:f^1_DR}) for
$f^{(1)}\equiv f_{\text{DR}}$ does not even require Hessians of $H$ or $\Delta
H$. Finally, we note that our result also agrees with a linearized
path-integral approximation obtained for a more general correlation function
$\operatorname{Tr}(Ae^{-iH^{\prime\prime}t/\hbar}Be^{iH^{\prime}t/\hbar})$ by
a similar approach by Shi and Geva \cite{Shi_Geva:2004a} in the context of
nonradiative electronic relaxation rates.

\emph{Example}: A sufficient condition for the first-order approximation
(\ref{eq:f^1_DR}) for fidelity amplitude to be exact is that the first-order
expansion (\ref{eq:DH_0}) itself is exact, which requires the average
Hamiltonian to be at most a quadratic function, and the perturbation at most a
linear function of positions and momenta, i.e.,%
\begin{align}
H  &  =\alpha+\beta\cdot q+\gamma\cdot p+q^{T}\cdot\delta\cdot q+p^{T}%
\cdot\varepsilon\cdot p,\\
\Delta H  &  =\Delta\alpha+\Delta\beta\cdot q+\Delta\gamma\cdot p,\nonumber
\end{align}
implying that the original Hamiltonians must be of the form
\begin{align}
H^{\prime}  &  =\alpha^{\prime}+\beta^{\prime}\cdot q+\gamma^{\prime}\cdot
p+q^{T}\cdot\delta\cdot q+p^{T}\cdot\varepsilon\cdot p,\label{eq:H_DR_exact}\\
H^{\prime\prime}  &  =\alpha^{\prime\prime}+\beta^{\prime\prime}\cdot
q+\gamma^{\prime\prime}\cdot p+q^{T}\cdot\delta\cdot q+p^{T}\cdot
\varepsilon\cdot p.\nonumber
\end{align}
In other words, the two Hamiltonians describe harmonic (or inverted harmonic)
systems that can be displaced in phase space, have different zeros of energy,
but must have the same \textquotedblleft masses\textquotedblright\ and force
constants in corresponding degrees of freedom. In one dimension, classical
motions corresponding to Hamiltonians $H^{\prime}$, $H^{\prime\prime}$ are
motions along ellipses or hyperbolas in phase space, where the centers of
these conical sections in phase space may be displaced between $H^{\prime}$
and $H^{\prime\prime}$, but otherwise the phase portraits must be the same for
the two Hamiltonians. For systems described by Hamiltonians
(\ref{eq:H_DR_exact}), the first-order approximation $f^{(1)}(t)$, i.e., the
dephasing representation, is exact for arbitrary initial states $\rho$. Such
systems can be used to describe, e.g., electronic absorption and emission
spectra in molecules, where the displacement occurs only in coordinate space
(i.e., $\Delta\beta\neq0$ and $\Delta\gamma=0$) and results in vibrational
excitation of a molecule upon electronic absorption. In contrast, Hamiltonians
with displacement in momentum space ($\Delta\beta=0$ and $\Delta\gamma\neq0$)
are useful for representing inelastic collisions, such as inelastic neutron
scattering \cite{Petitjean_Bevilaqua:2007}.

Indeed, it is not surprising that the first-order approximation
(\ref{eq:f^1_DR}) is exact for quadratic Hamiltonians with linear
perturbation, since many semiclassical approximations are exact in such
situations. What is intriguing about the dephasing representation
(\ref{eq:f^1_DR}) is its surprisingly accuracy in chaotic systems. So the
approximation is exact for Hamiltonians (\ref{eq:H_DR_exact}) and accurate in
chaotic Hamiltonians, yet the most severe breakdown for it occurs in simple
systems, such as quadratic Hamiltonians with quadratic perturbations. Next we
turn to deriving an expression that will correct this drawback.


\subsection{Second-order expansion}

In order to simplify the presentation of the second-order expansion, we shall
assume that $D=1$ and $\Delta H(x)\equiv\Delta V(q)$. The quadratic expansion
of (\ref{eq:DeltaH_expansion}) approximates the Hamiltonian difference as%
\begin{equation}
H^{\prime\prime}(x^{\prime\prime})-H^{\prime}(x^{\prime})\approx\Delta
V(q)+\frac{\partial T}{\partial p}\Delta p+\frac{\partial V}{\partial q}\Delta
q+\frac{1}{8}\frac{\partial^{2}\Delta V}{\partial q^{2}}(\Delta q)^{2}.
\label{eq:DH_2}%
\end{equation}
With this expansion, the phase (\ref{eq:A}) in the path integral
representation (\ref{eq:f_PI_2}) becomes%
\begin{equation}
A_{N}^{(2)}=-\tau\sum_{n=1}^{N}\Delta V(q_{n})+\sum_{j=0}^{N-1}\left(
q_{j+1}-q_{j}-\tau\frac{\partial T}{\partial p}(p_{j})\right)  \Delta
p_{j}-\hbar\sum_{k=1}^{N}\left(  a_{k}\left(  \Delta q_{k}\right)  ^{2}%
+b_{k}\Delta q_{k}\right)  -p_{0}\Delta q_{0}, \label{eq:A^(2)}%
\end{equation}
where%
\begin{align}
\hbar a_{k}  &  :=\frac{\tau}{8}\frac{\partial^{2}\Delta V}{\partial q^{2}%
}(q_{k})\text{ \ \ and}\\
\hbar b_{k}  &  :=p_{k}-p_{k-1}+\tau\frac{\partial V}{\partial q}%
(q_{k}).\nonumber
\end{align}
Again, the integrals over $\Delta p_{n}$ in (\ref{eq:f_PI_2}) yield delta
functions with arguments agreeing with Hamilton's equations of motion for
$q_{n}$, and the integral over $\Delta q_{0}$ gives the Wigner function of the
initial state:
\begin{align}
f^{(2)}(t)  &  =\int\frac{d^{2}x_{0}}{h}\rho_{W}\left(  x_{0}\right)
\prod_{n=1}^{N}\frac{d^{2}x_{n}}{h}d\Delta q_{n}\delta\left(  q_{n}%
-q_{n-1}-\tau\frac{\partial T}{\partial p}(p_{n-1})\right)  e^{iB_{N}%
^{(2)}/\hbar},\label{eq:f^(2)_1}\\
B_{N}^{(2)}  &  :=-\tau\sum_{n=1}^{N}\Delta V(q_{n})-\hbar\sum_{k=1}%
^{N}\left(  a_{k}\left(  \Delta q_{k}\right)  ^{2}+b_{k}\Delta q_{k}\right)
\nonumber
\end{align}
Although the complex Gaussian integrals over $\Delta q_{n}$ do not yield
simple Dirac delta functions, they can be evaluated analytically, and the
fidelity amplitude becomes%
\begin{align}
f^{(2)}(t)  &  =h^{-1}\int d^{2}x_{0}\rho_{W}\left(  x_{0}\right)  \prod
_{n=1}^{N}d^{2}x_{n}\delta\left(  q_{n}-q_{n-1}-\tau\frac{\partial T}{\partial
p}(p_{n-1})\right)  \tilde{\delta}\left(  p_{n}-p_{n-1};q_{n}\right)
\nonumber\\
&  ~~\times\exp\left[  -\frac{i}{\hbar}\tau\sum_{k=0}^{N-1}\Delta
V(q_{k})\right]  , \label{eq:f^(2)_2}%
\end{align}
where $\tilde{\delta}$ is a \textquotedblleft smeared\textquotedblright\ delta
function, given by a complex Gaussian
\begin{equation}
\tilde{\delta}\left(  p_{n}-p_{n-1};q_{n}\right)  :=h^{-1}\int d\xi
e^{i(a_{n}\xi^{2}+b_{n}\xi)}=h^{-1}\sqrt{\frac{\pi}{|a_{n}|}}\exp\left[
\frac{i}{4}\left(  \pi\operatorname{sgn}a_{n}-b_{n}^{2}/a_{n}\right)  \right]
.
\end{equation}
This smeared delta function replaces Hamilton's equation for $p_{n}$ with a
\textquotedblleft smeared Hamilton's equation\textquotedblright---the
expectation value of momentum $p_{n}$ is still at the classical value
$p_{n-1}-\tau\frac{\partial V}{\partial q}(q_{n})$, but it is not
deterministic as in classical mechanics. Equation (\ref{eq:f^(2)_2}) for the
second-order fidelity amplitude thus has a simple interpretation, not unlike
the dephasing representation: First, one samples initial conditions $x_{0}$
from the density $\rho_{W}(x_{0})$. Then one runs trajectories starting from
these points, where the kinetic propagation of positions is classical and
hence deterministic, whereas the propagation of momenta is nonclassical and
stochastic. Although we have been able to evaluate three quarters\ of the
integrals in the exact path integral representation (\ref{eq:f_PI_2}) of
$f(t)$, the remaining $N$ integrals over $p_{n}$ render the resulting
expression (\ref{eq:f^(2)_2})~still a formidable path integral, which is
difficult to evaluate numerically. Note that if we allowed the perturbation to
affect also the momenta, then the propagation of positions would also be
stochastic; the corresponding generalization of Eq.~(\ref{eq:f^(2)_2}) is straightforward.

\emph{Example}: A sufficient condition for the second-order approximation
(\ref{eq:f^(2)_2}) for fidelity amplitude to be exact is that the second-order
expansion (\ref{eq:DH_2}) itself be exact, which requires the average
Hamiltonian to be at most a quadratic function of $q$ and $p$, and the
perturbation at most a cubic function of $q$, i.e.,%
\begin{align}
H  &  =\alpha+\beta\,q+\gamma\,p+\delta\,q^{2}+\varepsilon\,p^{2},\\
\Delta H  &  =\Delta\alpha+\Delta\beta\,q+\Delta\delta\,q^{2}+\Delta
\phi\,q^{3},\nonumber
\end{align}
implying that the original Hamiltonians must be of the form
\begin{align}
H^{\prime}  &  =\alpha^{\prime}+\beta^{\prime}q+\gamma\,p+\delta^{\prime}%
q^{2}+\varepsilon\,p^{2}-\frac{1}{2}\Delta\phi\,q^{3},\\
H^{\prime\prime}  &  =\alpha^{\prime\prime}+\beta^{\prime\prime}%
q+\gamma\,p+\delta^{\prime\prime}q^{2}+\varepsilon\,p^{2}+\frac{1}{2}%
\Delta\phi\,q^{3}.\nonumber
\end{align}


\section{\label{sec:discussion} Discussion}

The derivations based on the Feynman path integral bypass the conventional
semiclassical approximations and therefore allow us to introduce several
rigorous statements. If the Hamiltonian is up to\emph{ quadratic} and the
perturbation up to\emph{ linear}, the dephasing representation (or phase
averaging \cite{Mukamel:1982}\ or weighted average classical limit
\cite{Egorov_Rabani:1998}) is exact. For example, for displaced simple
harmonic oscillators
\begin{align}
H^{\prime}  &  =\frac{p^{2}}{2m}+\frac{1}{2}k\left(  q-\frac{a}{2}\right)
^{2},\\
H^{\prime\prime}  &  =\frac{p^{2}}{2m}+\frac{1}{2}k\left(  q+\frac{a}%
{2}\right)  ^{2},\nonumber
\end{align}
the dephasing representation is exact \cite{Mukamel:1982} if the classical
trajectories are propagated with the average Hamiltonian $H$ since then the
Hamiltonian difference (\ref{eq:DeltaH_expansion}) is indeed linear in $\Delta
q$ and $\Delta p$:
\begin{equation}
H^{\prime\prime}(x^{\prime\prime})-H^{\prime}(x^{\prime})=\frac{p}{m}\Delta
p+kq(a+\Delta q).
\end{equation}
In contrast, the dephasing representation is \emph{not} exact even in this
simple system if $H^{\prime}$ is used for dynamics since quadratic terms in
both $\Delta q$ and $\Delta p$ appear:%
\begin{equation}
H^{\prime\prime}(x^{\prime\prime})-H^{\prime}(x^{\prime})=\frac{1}%
{2m}(2p^{\prime}+\Delta p)\Delta p+\frac{1}{2}k(2q^{\prime}+\Delta q)(a+\Delta
q). \label{eq:expansion_H_about_H'}%
\end{equation}
Similarly, the dephasing representation is \emph{not} exact (in fact, breaks
down rather severely) for simple harmonic oscillators with different force
constants,%
\begin{equation}
H^{\prime}=\frac{p^{2}}{2m}+\frac{1}{2}k^{\prime}q^{2}\text{ \ and }%
H^{\prime\prime}=\frac{p^{2}}{2m}+\frac{1}{2}k^{\prime\prime}q^{2},
\end{equation}
since the perturbation is quadratic in $\Delta q$ even if the average
Hamiltonian is used for dynamics:%
\begin{equation}
H^{\prime\prime}(x^{\prime\prime})-H^{\prime}(x^{\prime})=\frac{p}{m}\Delta
p+kq\Delta q+\frac{1}{2}\Delta k\left[  q^{2}+\frac{1}{4}\left(  \Delta
q\right)  ^{2}\right]  .
\end{equation}
The last simple example provides a particularly bad scenario for the dephasing
representation, which can be remarkably accurate in much more complex, even
chaotic systems such as the kicked rotor. Unfortunately, undisplaced harmonic
oscillators provide a good model for the \textquotedblleft
silent\textquotedblright\ modes in electronic spectra, i.e., the modes which
are not excited by the electronic transition, and hence are not displaced, but
may have a different force constant in the excited state. Especially in large
molecules, the majority of the modes are silent, but the dephasing
representation produces an artificially fast decay of fidelity amplitude
\cite{Zambrano_Vanicek:2013}, which in turn gives rise to artificially
broadened spectra, often to the point that any structure is lost. Typical
molecules are slightly anharmonic, so one cannot always use simple
semiclassical methods such as the thawed Gaussian approximation
\cite{Heller:1975}, but they are not very chaotic, and hence the surprising
accuracy of dephasing representation in chaotic systems does not help. Yet,
the second-order approximation (\ref{eq:f^(2)_2}) for $f(t)$, which is, by
definition, exact in harmonic systems with different force constants,
could---if evaluated efficiently---provide an accurate method for computing
electronic molecular spectra even in the presence of anharmonicity and
wavepacket splitting.


\section{\label{sec:conclusion}Conclusions}

In conclusion, we derived a path integral formula for the quantum fidelity
amplitude, which bypasses the conventional semiclassical approximations of
past publications. Our first approach used a coordinate path integral for
continuous systems and benefited from the explicit connection with the
classical Liouville propagator. We note that this path integral approach
allows in principle to incorporate the influence of the environment using the
familiar Feynman-Vernon formalism. All that is required is adding the
appropriate bath terms to the action. The effect of thermal noise would be to
broaden the delta functions that arise from the $\mathcal{D}r$ integration,
leading to smearing of the phase-factor in Eq.~(\ref{eq:ft}).

Our second approach relied on the phase-space path integral for
\rmrk{kicked} quantum maps. In the latter context we also obtained
an exact expansion of the exponent of the path integral and derived explicit
expressions for the fidelity amplitude in the zeroth, first, and second-order
expansions; the first-order expansion yields exactly the dephasing
representation, whereas the second-order expansion yields an approximation
which corrects several drawbacks of the dephasing representation and other
approximations based on linearizing the semiclassical propagator or path
integral. It remains to be seen if it can be implemented efficiently.

Finally, the rigorous manipulation of the path integral has allowed us to make
several rigorous statements about the validity of various approximations for
fidelity amplitude.

\emph{Acknowledgments}. We would like to thank Marius Wehrle for discussions.
This research has been supported by the Swiss National Science Foundation
(grant No. 2000201\_50098) and by the Israel Science Foundation (grant No. 29/11).



%


\end{document}